\documentclass{acm_proc_article-sp}

\begin{document}

\title{ Distributed Air Traffic Control : A Human Safety Perspective}

\numberofauthors{3} 
%
\author{
%
%
\alignauthor
Sarvesh Nikumbh\\
       \affaddr{Centre for Modeling and Simulation}\\
       \affaddr{University of Pune}\\
       \affaddr{Pune 411007, India}\\
       \email{sarvesh@cms.unipune.ac.in}
 \alignauthor
 Joeprakash Nathaman
        \email{joeprakash.n@gmail.com}
 \alignauthor 
 Rahul Vartak
        \email{rahulyvartak@gmail.com}
}

\date{16 August 2011}

\maketitle
\begin{abstract}
The issues in air traffic control have so far been addressed with the intent to improve resource utilization and achieve an optimized solution with respect to fuel comsumption of aircrafts, efficient usage of the available airspace with minimal congestion related losses under various dynamic constraints. So the focus has almost always been more on smarter management of traffic to increase profits while human safety, though achieved in the process, we believe, has remained less seriously attended. 
This has become all the more important given that we have overburdened and overstressed air traffic controllers managing hundreds of airports and thousands of aircrafts per day.

We propose a multiagent system based distributed approach to handle air traffic ensuring complete human (passenger) safety without removing any humans (ground controllers) from the loop thereby also retaining the earlier advantages in the new solution. The detailed design of the agent system,  which will be easily interfacable with the existing environment, is described. Based on our initial findings from simulations, we strongly believe the system to be capable of handling the nuances involved, to be extendable and customizable at any later point in time.

\end{abstract}


\section{Introduction}
Air traffic is increasing rapidly in many developing and developed countries. The busiest airports are swamped everyday and incidents involving overworked and overstressed air traffic controllers (ATCs) is increasing. In the current system, ATCs have some awareness of things related to major airport closures, but they have to make multiple decisions about how to move and re-route traffic just based on their experience and little help to guide them. That's better than nothing, when there are about 16000 ATCs at more than 5000 public airports in the USA itself and more than 13000 worldwide \cite{tumer:atm}. The end result is decisions that are not always optimal for the larger, national needs. The Controllers are on the job for more than 16 to 17 hours on the trot. The resulting fatigue and stress invariably transfers on their decision making, which may result in fatal or penalizing errors. As a matter of fact, there have been many such incidences in the past couple of years. It is this aspect of transferred risk of safety of passengers due to overburdened controllers or humans is what we try to address in this work while the usual optimization of resources is inherently achieved.

\section{The {\secit Multiagent} System}
In this work we propose a multiagent system based distributed approach for such modern air traffic controlling. Modeled as an inherently fault tolerant system, it handles efficient holding pattern and landing sequence generation. It provides a solution out of the classical centralized routing strategies which are often slow to respond to developing weather conditions or airport operations, and which allow minor local delays to cascade into large regional congestions. We restrict our approach to operate only in the vicinity of airports. In this work, we attempt to address the issues of : 
\begin{itemize}
 \item What constitutes an agent in this system?
 \item What actions can the agents take to impact air traffic?
 \item Its interface with the environment.
\end{itemize}
Focussing on the airport vicinities lets us proceed with the obvious choice of an aircraft being an agent though in literature there are instances of \textit{fixes} \textendash which indicate ground locations throughout the airspace \textendash as choice of agents. This choice of aircrafts as agents helps us achieve the aim of being able to tackle issues arising in-flight, on-board that trigger last minute decision changes \cite{avweb:aviation-news}. 

\subsection{The Design}
Each aircraft is responsible with various tasks according to the phase of flight it is currently in. The various phases of flight are viz. arrival into airport vicinity (henceforth called the airspace), at entry gate, on path, holding pattern, to metering fix, at metering fix, final descent, on runway and backtrack. Their service requirement is then dynamically determined by arising circumstances. On arrival into airspace, every aircraft shares a copy of queue Qi that has a buffer of size Bi to accommodate other aircrafts (whose number is limited for this airspace). We have holding patterns formed which are stacks of aircrafts hovering in the airspace at different altitudes maintaining the queue. These aircrafts in the stack form sets of agent systems. All of those at nearly the same altitude (or groupable altitudes) form a system \textit{\textquoteleft A\textquoteright} which is lead by the first amongst them in the queue to land. Similarly there will be systems \textit{\textquoteleft B\textquoteright}, \textit{\textquoteleft C\textquoteright} and so on at higher altitudes lead by the corresponding leaders. The leaders in all the stages also form a system amongst themselves together with the aircrafts which have just entered into flight on takeoff. This helps them be updated with the direct information of the aircraft's flight just upon entering the airspace. The non-leaders in each stage would have this information relayed to themselves through the leaders. Once the leader at the lowest altitude takes to landing and leaves the system it passes on the leadership to the next in line. Each aircraft here, on entering into any agent system, is updated with the corresponding current status by a provisional agent \textendash \space the run time \textit{Directory Facilitator (DF)}. It is responsible for providing every agent in the system with the required information about other agents currently alive and their co-ordinates, state and phase description. The ATC isn't thrown out-of-the-loop and is infact our next agent in the system. It picks up major properties as mentioned in \cite{lotto:francois}. It overhears and overviews completely the decisions made in the entire process. ATC is the main passive supervision service provider requiring very little amount of service for itself. It shares the queue \textit{$Q_{i}$} with buffer of size \textit{$B_{i}$} to track all the aircrafts and their movements. It is also responsible for handling communication with other running processes such as the Surface Movement Radars (SMR), TRACONs (Terminal Radar Approach CONtrol). It does so with the help of the other provisional agent of the system \textendash \space \textit{InProcess interfacing} agent. It is responsible for detecting agents in the environment and continually updates the \textit{DF} agent with this piece of information. It reads input from other external processes or entities on the traffic control system such as Flight Management System (FMS), User Request Evaluation Tool (URET) and many others. The other important agent of the system is the TRACON. It is the second control facility located within the vicinity of a large airport typically controling aircraft within a 30-50 nautical mile radius of the airport between the surface and 18,000 feet. The actual airspace boundaries and altitudes assigned to a terminal control are based on factors such as traffic flows, neighbouring airports and terrain, and vary widely from airport to airport. It is entrusted with ATCs' responsibilities in their absence. They also notify the ATC regarding any aircraft entering the concerned airspace. They thus form the fall-back agents offering a stronger fault-tolerance. We can also handle communication with \textit{\textquoteleft fixes\textquoteright}, the agents for the greater airspace facilitating seamless interface between other agent systems elsewhere in the environment. This way we are guaranteed to gain from the advantages of both \textendash \space \textit{fixes} and \textit{aircrafts} \space \textendash \space as agents. Each agent can now effect a re-route or a ground delay or cause a change in the sequence by prompt  decision making. It prevents linguistic/mental communication errors triggered by humans in the loop. These \textquoteleft \textit{well-aware}\textquoteright aircrafts would thus be in a position to tackle issues of emergency re-routes or re-sequencing without any fatal delay.

\section{Simulations}
We based our simulations on artificial data with typical conditions viz. free-flights maintained as per the path-corridor correlation \cite{pendse:cde} and the data-exchange between agents follows the co-operative data exchange described in \cite{charter:aircraft}. The combinations of dynamic causes for route or flight-plan changes in the participating agents were chosen by Monte Carlo methods \cite{archibald:architecture}. The resulting variations observed in the pre-fixed flight paths were succesfully handled under various scenarios. The simulations were conducted for Mumbai airport which has two runways (landing and takeoff) crossing each other. We achieved an average of about 38 flights per hour with average successful handling of 86.67 \% of all flights with variations in their flight-paths, without human intervention. The results are averaged over 25 runs.

\section{Future Work}
There have been issues observed previously in the Indian airspace caused due to some weak links in the administration. For instance, a well-known public carrier, in 2008, was to land at Hyderabad airport but was allegedly unaware of the new Shamshabad airport inaugurated only a few days before, leading the confused pilots to fly the aircraft to Delhi and then Mumbai (reported in the News \cite{avweb:aviation-news} and also documented on Wikipedia). Such incidences can be avoided given that a secure information-exchange protocol can ensure relaying of important notifications across aircrafts. Also, once the system gets into operation, over time it could learn by itself the various traffic patterns observed over a single day for an airspace and bring about an improvement in the overall performance thus optimizing usage of invaluable resources. Also we could have more well defined reward mechanisms \cite{tumer:atm} implemented for the agents improving their performance as time progresses. 

\section{Conclusion}
A fundamental element of this research has been to investigate the air traffic operational environments and the effect of and on all the humans in the loop. The need for automating certain tasks and still leaving the monitoring part to the ground air traffic  controller has been made visible. The multiagent system and its internal details are discussed. The simulations are performed on the basis of work presented previously in literature. We do not expect the system to yield readily acceptable results in any given scenario. It needs to be rigorously tested to maintain its performance graph. Despite these limitations, we believe that our proposition and its demonstration clearly and quite precisely explain the handling of all the nuances in the system and are customizable with relative ease of intuition. We are confident that it can be extended to a variety of similar real-world problems.

\bibliographystyle{abbrv}
\bibliography{Sarvesh_Nikumbh}

\end{document}